\documentclass[preprint]{revtex4-1}
\begin{document}

\title 
      {Lessons in quantum gravity from quantum field theory}

\pacs{11.25.Tq, 11.15.Pg}
\keywords{AdS/CFT correspondence, emergent geometry}

\author{David Berenstein}

\address{Department of Physics, University of California at Santa Barbara, CA 93106\\ Institute for Advanced Study, School of Natural Science, Princeton, NJ 08540}

\begin{abstract}
This paper reviews advances in the understanding of quantum gravity based on field theory calculations in the AdS/CFT correspondence.
\vfil

{\em Presented for the Proceedings of the Symposium on Gravitation and BECÕs Phenomenology from the Fourth Mexican Meeting on Mathematical and Experimental Physics.}
\end{abstract}

\date{\today}

\maketitle

\section{Introduction}

It is an undeniable fact that the gauge/gravity duality \cite{M} has revolutionized the way we think both about gauge field theories and about quantum gravity. In many senses, the gauge/gravity duality is the first non-perturbative definition of quantum gravity on some classes of spacetimes that is consistent as a quantum system and for dimensions greater than two, the critical dimension for perturbative gravity.

The right place to begin is by asking what is a theory of quantum gravity? The standard answer is that it must be a theoretical framework where the following conditions are satisfied: first, the theory must allow us to describe scattering processes at ultrahigh energies, beyond the Planck scale and in the strong gravity regime, at energies where small black holes start making their appearance. We expect it to be consistent with the rules of quantum mechanics, so that unitarity is preserved. This is not guaranteed, as  Hawking famously argued that black holes lose information \cite{Hawking}. Secondly, such a theory must be compatible with everything that we know about gravity and particle physics at low energies: physics should be described by a classical geometry in this regime, a geometry that satisfies Einstein's equations and where particle quanta scatter as described by local quantum field theory calculations. Another condition that seems to be required is that the theory should be holographic \cite{T'HS}. 

The gauge /gravity duality and more precisely, the original AdS/CFT correspondence of Maldacena \cite{M} sates that the maximally supersymmetric quantum field theory in four dimensions with gauge group $U(N)$ is exactly identical as a quantum system to type IIB string theory as a theory of quantum gravity (that includes string branes and changes of topology) on spacetimes whose asymptotic geometry is $AdS_5\times S^5$ with some additional flux conditions.

The conjecture originated from the study of D-brane dynamics in string theory. D-branes are topological defects where strings can end. Their fluctuations are described by open string dynamics. If we go to the low energy regime, we end up getting only the shortest strings: they are massless or nearly massless. The infrared dynamics consists of particles with spin less than or equal to one: we get an ordinary field theory on the worldvolume of the brane. In the special case
studied in \cite{M} the objects under study where D3-branes in type IIB string theory. These preserve half of the supersymmetries, and the wolrvolume theory one obtains is ${\cal N}=4 SYM$ 
for the group $U(N)$. The branes are topological defects in a theory of gravity, they have a tension, so we can also study how they bend spacetime. Solving Einstein's equations lead to a region with an infinite redshift, similar to a black hole horizon. In this region, even heavy strings are light due to redshift effects. Thus any object placed in the near horizon geometry is also an infrared excitation. The claim stated in \cite{M} is that we can not tell apart these two ways of thinking about the infrared physics: one is more appropriate for computations when the branes cause little back reaction (weak coupling), and the other is more appropriate when the branes deform the geometry substantially (strong coupling).

What is truly remarkable is that this is a definition of quantum gravity that seems to satisfy all of the conditions required for a theory of quantum gravity and if true, it encodes ten dimensional dynamics in a four dimensional field theory. If we introduce the variable $R$ to described the size of the $AdS$ geometry in string theory units, we find that up to normalization constants $g^2_{YM}N \simeq R^4$. We also have that the closed string coupling constant scales as $g_s \simeq g_{YM}^2$.

The classical gravity regime occurs when $R$ becomes very large and $g_s$ is small. This requires us to take $N\to \infty$ and the t' Hooft coupling $g^2_{YM}N =R^4$ also becomes large, while keeping the coupling constant $g_s$ fixed and very small. From the field theory point of view, the ten dimensional dynamics as a field theory phenomenon requires a reorganization of the degrees of freedom of the field theory. The extra dimensions that appear in gravity are to be considered as a description of emergent phenomena in the usual sense of condensed matter physics: they require some collective behavior of the field theory variables at strong coupling that is not apparent on the standard perturbative description. 

This paper will describe various field theory computations that have been done not only to check the correspondence, but that I argue contain deep information about quantum gravity and the origin of geometry. Space limitations force me to leave out many other advances that I also consider very interesting. I hope that the list of references can serve as a starting point for a more serious study of these topics, but I will not claim it to be comprehensive. Although I have tried to keep the attribution straight to the original sources, I might have missed a few. To those left out I apologize.

\section{Observables}

Evidence in favor or against  the AdS/CFT conjecture requires comparing computations done in field theory and in the gravity formalism. We also have to be able to be in a regime where we can trust both computations simultaneously: this is usually where computations fail. Our techniques for computing many times have no overlapping regimes of validity.

The first place to begin is by understanding how to calculate quantities both in field theory and in the (quantum) gravity theory. 

The ${\cal N}=4 $ SYM is a superconformal field theory. This implies that apart from the usual Lorentz  symmetries (or supersymmetries) , we also have scale invariance and conformal transformations as sysmmetries. All together, the theory must have a symmetry $PSU(2,2|4)$, this includes the supersymetry. This symmetry acts by unitary transformations on a Hilbert space. Whatever this Hilbert space is, it must be the same in the field theory and the gravity side, and both must share the same representation theory under the $PSU(2,2|4)$ symmetry. This representation theory can be computed perturbatively in the weak gravity limit, but we have a hard time describing it in situations in which gravity is strong (black holes, etc). 

The ${\cal N}= 4 $  SYM is conformal. At the conformal point everything is massless, so the concept of particles is somewhat ill-defined, as one can add soft radiation to any process and one needs to take care to ensure that whatever one is computing is infrared safe. It is convenient to Wick rotate the theory to a Euclidean setup. In such a setup, the correlation functions of local observables give
a useful set of observables to study. This is, we consider gauge invariant local operators ${\cal O}_i(x_i)$ and we study their expectation values as functions of the $x_i$. We get that the observables are given as
\begin{equation}
G(x_1, x_2, \dots x_i)= \langle {\cal O}_1(x_1){\cal O}_2(x_2)  \dots {\cal O}_i(x_i)\rangle
\end{equation}
The positions of the $x_i$ serve to introduce a scale in the problem and regulate the infrared divergences.
 For massive theories, one can also take limits of these correlations and after Wick rotating to Lorentzian space, we end up with a description that permits one to calculate the S-matrix.
 The $G$ green's functions define the theory.

One can also regulate the theory by putting it on a finite size box and evolve it in time. Because of the operator-state correspondence of Conformal Field theories, the ideal box is a sphere. Thus for the ${\cal N}=4 $ SYM in four dimensions, one should place it on $S^3\times {\bf R}$, where the time is a real time. In such a setup, the moduli directions are lifted because the scalar fields have a conformal coupling to the metric and the spectrum is discrete. This spectrum ends up being in one to one correspondence with the local operators ${\cal O}(0)$. The energy in the sphere is identical to the scaling dimensions of the operators ${\cal O}(0)$ under dilations: we end up diagonalizing the dilatation operator rather than the momentum.  This is natural if one considers that the origin is fixed by dilatations.
The $G$ Greens functions can be recovered from this formulation.

For the gravity side the problem requires some more work. The isometry group of the $AdS_5\times S^5$ space is also $PSU(2,2|4)$. It is important to notice that these diffeomorphisms are not gauged: the $AdS_5\times S^5$ is non compact. These global rotations have meaning and their generators can be evaluated on configurations as ADM charges.

Again, it is convenient to go to the Eulcidean formulation and go to the Wick rotated version of $AdS_5$. This Euclidean $AdS_5$ is nothing other than the Poincare upper half plane in five dimensions: the maximally symmetric space with negative curvature. This space is non-compact, and has an asymptotic boundary that can be rescaled to finite distance by a Weyl rescaling. 

One of the properties of gravity  in such a Euclidean setup is that it has no normalizable perturbations. As such, our usual rules to study path integrals around  families of classical solutions of the theory would allow only one geometry: $AdS_5$ itself, and the gravity observable would be the gravity partition function: it is a number. To allow for more saddle points (more classical solutions), we should relax the requirement for normalizability. We should allow non-normalizable perturbations. These must be singular near the boundary of the Euclidean space. The perturbations would be characterized by some boundary data $\delta \lambda (x_i)$ which we can vary, and then we have one observable: the partition function of gravity with prescribed boundary data. The $\delta \lambda$ indicate the collections of all possible fields in the gravity theory.  The boundary of $AdS_5$ is four dimensional, so we find that a natural observable is
 \begin{equation}
 Z_{grav} = Z|_{\delta \lambda(x_i) fixed}
 \end{equation}
which can be evaluated as a saddle point around a classical solution with prescribed $\delta \lambda$. Now, all we need to be able to do is to compare the two ways of obtaining observables: the one in gravity and the one in field theory.
 The fact that the boundary data is four dimensional and characterized by functions that depend on $x_i$ suggest that the gravity fields and the list of operators can be related somehow: they are parametrized by similar data (positions), and there is a long list of gravity fields, while there is also a long list of local operators. 
 
The dictionary relating these two was established  by Gubser, Klebanov, Polyakov and Witten in \cite{GKPW}. The gravity field boundary conditions act as sources of the operators for a generating series of correlation functions. The master equation is 
\begin{equation}
\left \langle \exp  \left[\int d^4 x \sum_i \delta\lambda_i(x) {\cal O}_i(x)\right]  \right\rangle =
Z|_{grav, \delta \lambda(x_i) fixed}
\end{equation}
where the ${\cal O}_i$ are single trace operators. 

In a certain sense this dictionary compares two black boxes to check if they agree. Their agreement for general calculations would be the stated equivalence between the two theories. The invariance under Weyl rescalings of the theory tells us that the mass of the field associated to the perturbation $\delta\lambda$ is related to the dimension of the operators ${\cal O}_i$. 

Taking derivatives with respect to the $\lambda$  at $\lambda=0$ can  be used to recover the $G$ for general correlators. Also, one can recover condensates at finite values of $\delta \lambda$ by taking derivatives of the expression if the operators ${\cal O}$ are relevant. In the gravity theory this involves studying the subleading asymptotics of the gravity fields.

Let me explain how this can be used to explore properties of the field theory at strong coupling. Let us say that we want to study field theory at finite temperature. It is argued that this is equivalent to working with AdS black hole solutions of the gravity theory \cite{Witten}, after all, the conformal boundary of the Euclidean black hole solution is equivalent to $S^3\times S^1$, where the periodic imaginary time gives us a temperature. To add chemical potentials, etc we source electric and magnetic fields in the black hole solution.

Given such a finite temperature problem, the field theory might look like some strongly coupled plasma. We might want to know transport coefficients of the field theory. These are notoriously hard to compute in perturbation theory. Not so in gravity. The field theory transport coefficients are computed by the Kubo formula: these involve two point function correlators of the stress tensor and the currents. In response theory one should use retarded propagators to define these dynamical quantities.

Given the classical black hole geometry, these retarded Green's functions are classical Green's functions on the black hole geometry. The boundary conditions are infalling into the black hole. This permits one to study viscosities, sound, conductivities, etc, by performing a classical gravity computation. This has been applied to the problem of the RHIC fireballs \cite{PSSK} and problems in condensed matter physics, especially superconductivity \cite{Hetc}. This is well reviewed in \cite{Hartrev} and it has led to an enormous list of solvable models of strong coupling setups, although the corresponding field theories tend to be at large N and sometimes we do not know the microscopic theory (the gauge field theory), but only the gravity backgrounds. These details are beyond the topic of this presentation. What we will explore in the rest of the paper is the harder step of trying to deduce information about gravity from attempting to solve the field theory at strong coupling.

\section{Perturbation theory and integrability}

As described before, the spectrum of operators of the field theory and the representation theory of these under the symmetry of the ${\cal N}=4 $ SYM is important. The spectrum of scaling dimensions of operators can be computed in perturbation theory and one can start to work a systematic theory of how these are organized. The first breakthrough in this direction occurred in \cite{BMN} where it was shown that the field theory not only included operators associated to massless gravity modes, but that one could also show that the list of operators for large quantum numbers also encoded the spectrum of strings in a very particular geometric limit and that the calculations could be matched: the string spectrum could be compared directly with the field theory.
The problem has very many operators that are degenerate in the free field limit. One then has a problem in degenerate perturbation theory. In the planar limit, the spectrum was made to look like a gas of impurities on a large lattice  with nearest neighbor hopping (and at higher orders next to nearest neighbor hopping, etc), so that one could compute a dispersion relation for the impurities at various orders in perturbation theory and eventually to all orders \cite{SZ}. This result gave very strong evidence for the AdS/CFT correspondence. 

It was followed by the work \cite{MZBS} where it was shown that the planar perturbation theory theory could be organized as a spin chain model for an integrable system. Afterwards it was shown that the sigma model for the string on $AdS_5\times S^5$ was also integrable  not only classically but also at the leading quantum level \cite{BPR}. This led to the conjecture that the strong and weak coupling limit of the theory are related by a one parameter family of quantum  integrable models \cite{DNW}. The evidence for this particular proposal is overwhelming. The integrable model can be analyzed by guessing a Bethe Ansatz that can be matched both to perturbation theory and to the supergravity regime. The interpolating ansatz has been tuned after many papers and it has made remarkable predictions that have ben checked in perturbation theory to high loop order. The most recent proposal for a full solution of the planar system is given in \cite{GKV}. A lot of these developments have been reviewed in \cite{Serban}.

So now we are in a position to ask:  what does this mean for the program of string theory? The planar solution in terms of an integrable system means that we can in some sense solve exactly the theory of free strings moving on $AdS_5\times S^5$. This can be organized as a series in $\alpha' \simeq (g_{YM}^2N)^{-1/2}$ and given a solution for the string motion we can in principle read how the corrections in $\alpha'$ work in detail, although comparing to string theory $\alpha'$ corrections is hard. First of all, integrability and supersymmetry together produce a huge set of cancellations between various contributions,  but we only  get the final answer and no hint at how all the various different terms of a worldsheet perturbation expansion contribute to it.
 This is strictly at $N=\infty$, so the exact solution of the string on the geometric background can be done with stringy quantum corrections included. Gravitational corrections involve the non-planar corrections to the power series in $1/N^2$ and this is not understood in this formulation. In particular, the problem of  how strings source a gravitational field is encoded in these corrections. This means that we have not been able to reproduce this basic aspect of gravity yet. 
Notice also that the assumption of integrability is very special. Most string backgrounds are not solvable this way.

Surprisingly, there are new backgrounds associated to field theories in three dimensions that seem to share this property \cite{ABJM}, a calculation that was done in \cite{MZ2}. Since these examples involve different gravitational  background fields, we get other information on $\alpha'$ corrections this way. However, in this case there is an interpolating function called $h(\lambda)$ that is not yet known that is required in order to make the strong and weak coupling Bethe Ansatz match at the level of the dispersion relation for magnons which is part of the input data for the Bethe Ansatz.

\section{Exact results for Wilson loops}

A second class of observables in gauge theory are Wilson loops. These are not local (point like) insertions in the path integral, but instead they involve integrating fields over a closed path. These can be used to describe the quark-anti quark potential and also serve as an order parameter for confinement. These Wilson loops are usually written as
\begin{equation}
W({\cal C}) = \left \langle \hbox{Tr }_R  P\left(\exp( i \oint_{\cal C} A_\mu dx^\mu) \right) \right \rangle
\end{equation}
where the $P$ indicates path ordering the exponential, and $R$ is a representation of $U(N)$. In the simplest case we take the fundamental.

In the case of the Maldacena conjecture, the Wilson loops result in insertions of a heavy quark circulating on the geometric loop. In the gravity dual, the fundamental strings ending on the branes are the objects that carry the fundamental charge under the gauge group. Thus we should work with a classical string whose end-points lie on the Wilson loop curve on the boundary of $AdS$. Because these quarks also coupled to the scalars of the SYM theory, the Wilson loop that is calculated in gravity is given by 
\begin{equation}
W({\cal C}) = \left \langle \hbox{Tr }_R  P\left(\exp( i \oint_{\cal C} A_\mu dx^\mu +\phi_6 ds) \right) \right \rangle
\end{equation}
where $\phi_6$ is one of the scalars in the SYM multiplet. In the gravity theory one calculates the action of a classical world sheet that ends on the curve ${\cal C}$ \cite{MWil}.  For the circular loop, these operators preserve some of the supersymmetries and therefore it becomes plausible that  one might be able  to calculate them to all orders in field theory. A proposal for the answer was given in \cite{ESZ} plus various arguments for its validity. A rigorous proof appeared later in the work of Pestun \cite{Pestun} where it was shown that the field theory could be topologically twisted and therefore the partition function calculated at weak coupling can become an exact answer.

These results do sum not only to all orders in $\alpha'$, but also to all orders in $1/N$. The end result can be written in terms of a matrix integral. These results are very remarkable: they interpolate between the weak coupling perturbative series and the strong coupling behavior. The planar result for the fundamental representation is given by
\begin{equation}
W_\circ  = \frac 2{\sqrt{g^2N}} I_1(\sqrt{g^2 N})  \simeq \exp( \sqrt \lambda)
\end{equation}
where $I_1$ is a Bessel function. 

These expressions are remarkable in that they encode information that is valid for any value of the coupling constant and also in the region where gravity is not classical. There is a wealth of information waiting to be extracted on what these expressions mean for the theory of quantum gravity. Not only are these strong checks of the AdS/CFT duality, but in some sense they are telling what is the correct answer for gravitational corrections. 

Surprisingly, these techniques have also been useful in the three dimensional ABJM theory \cite{ABJM}. The Wilson loops have slightly different structure \cite{BT,Wall}. The proof that these computations are subject to solving them by localization techniques was done in \cite{Kapustin}, where the problem was reduce to a multivariable integral. It was shown later that this  integral is a solvable supermatrix model and that the Wilson loops that preserve half the supersymmetries also involve the fermion fields of the theory \cite{DruT}. These computations give us some information that is very similar to this unknown $h(\lambda)$ required for the Bethe Ansatz of the three dimensional planar theory. The calculations match both the free field perturbation theory and the supergravity regime. Remarkably, this is a setup where the geometry receives $\alpha'$ corrections (at least the string motion on it does) and we can match the two expansions. This case has less supersymmetry than the four dimensional AdS/CFT for maximally supersymmetric Yang Mills theory.

\section{Approximations at strong coupling and emergent geometry}

So far I have discussed calculations that have a geometric interpretation in gravity, but that from the field theory point of view all hints of the extra dimensional geometry are hidden.  The field theory output  gives us a function of the planar coupling constant, and sometimes also a full expansion in $1/N$. This is, we get a list of stringy corrections and quantum corrections in the gravity side that is very hard to interpret geometrically.  Strictly speaking, in  the thermodynamic limit of the spin chain model one can recover something that looks like a sigma model and match it to a certain limit of the string sigma model \cite{Kruc}, so one starts seeing a hint of geometry, but in a very quantum sense. One could imagine that it is possible to find the geometry of gravity in the field theory in a different way that makes it more apparent.

To get there, we need to do make a small detour into the classical theory of ${\cal N}=4 $ SYM on the space $S^3\times {\bf R}$ at weak coupling, but we will analyze it for large fields. The idea is simple: in the gravity side we not only have strings but also D-branes and other extended objects. In the large radius limit these are perfectly well described by classical configurations of branes probing the geometry, where we only need to solve the classical equations of motion with certain quantum numbers fixed.  If we need to we can  perhaps perform a semiclassical quantization of these solutions.  The simplest such solutions are giant gravitons \cite{McGreev}. These sates preserve half of the supersymmetries of the quantum gravity theory (they are half BPS). Their complete dual description can be done by  operators in the field theory build by combinatorial methods using Young Tableaux \cite{CJR}. 

However, we can ask the same classical questions for the ${\cal N}=4 $ SYM on the field theory side. What happens if we solve the classical equation of motion of SYM on the field theory side with certain quantum numbers fixed? Maybe some of these solutions persist to some degree of approximation even when we take the theory to strong coupling, especially so for states that preserve some supersymmetries. If one solves the half BPS equations on the sphere one finds that the classical solutions are very simple: only the s-wave modes of the scalar fields on the sphere get  excited. Indeed, the solution is given by having only one scalar turned on $Z(t) = Z(0) \exp(it)$ where $Z$ is a complex scalar in the adjoint of $U(N)$. It is also necessarily a normal matrix. One can also show that restricting to these solutions the adjoint $\bar Z$ is canonically conjugate to $Z$.  Because of gauge invariance we can pass to the eigenvalues of $Z$ as the only dynamical degrees of freedom. The end system gives us $N$ free fermions on the a two dimensional phase space \cite{Btoy}. This is the same description as the integer quantum hall effect for the lowest Landau level of a system of electrons in a magnetic field. The electron gas forms an incompressible fluid and the ground state is a circular droplet. Fluctuations in the change of shape of the droplet describe
the excitations of the system. One can also add holes and move electrons away from the droplet and construct multi-droplet states. Each of these droplets then has a quantized number of particles. These are mapped to the giant gravitons once one understands how the eigenvalues are related to the combinatorics of  Young Tableaux. This description of these BPS states is valid at zero coupling, but the solutions of the equations of motion obtained this way are valid at finite coupling as well. The  droplet becomes macroscopic because $N$ is large and the Fermi statistics of the fermions forbid them to be piled on top of each other: this is a quantum effect. The origin lies in the fact that the system is gauged, the volume of the gauge orbit plays a very important role determining the effective repulsion of the eigenvalues.
 Surprisingly, a similar solution of all the half BPS geometries was found in the gravity theory where the description of the geometry is in terms of an incompressible fluid on a plane. This data provides a boundary condition for an elliptic differential equation in the gravity side. The data describes a degeneration locus of some fibration. The two coloring of the plane into empty and full regions is related to regularity of the supergravity solution \cite{LLM}. 

A  general class of all classical solutions of the BPS equations that preserve only $1/8$ of the supersymmetries is given in \cite{BLargeN}. These are again given by very simple solutions $X(t) = X(0)\exp(it)$, $Y(t)=Y(0)\exp(it)$, $Z(t) = Z(0)\exp(it)$, where $X,Y,Z$ are complex scalrs, constant on the sphere. In these solutions the $X,Y,Z$ are normal, and most importantly, they commute with each other. Again, just like in the case of the half BPS states, their adjoints are also the canonically conjugate variables to $X, Y, Z$. Going to eigenvalues (as allowed by gauge invariance), we find that we get $N$ bosons on ${\bf C}^3$. This is a quantization of the moduli space of the field theory. In order to make progress one wants to write wave functions in the moduli space, just as one does for the free fermions. There is an obvious polarization of this phase space, where one only uses holomorphic polynomials to describe wave functions. To calculate further one needs to know a measure 
to be able to compute averages in the quantum theory. This can be done strictly semiclassically: each eigenvalue represents three harmonic oscillators in the complex representation. We know the ground state classical wavefunctions of the harmonic oscillator and we multiply these. To include the fact that we passed to eigenvalues into the description,  we find that there is an induced repulsion between eigenvalues from the volume of the gauge orbit of a configuration. This is similar in nature to the fermion repulsion for the case of only one matrix. The wave function describes a gas of eigenvalues in six dimensions with long range repulsions and a confining external potential.
In the thermodynamic limit the eigenvalues are described by a density distribution on ${\bf R}^6\simeq {\bf C}^3$. All the eigenvalues are at the same distance from the origin and the density of eigenvalues is constant in the angular directions. We find a shape of an 5-sphere, $S^5$ this way. But the dual of this theory is supposed to be $AdS_5\times S^5$, which also has a 5-sphere in it. 

If one counts the fluctuation spectra, we end up with the same results and this suggest very strongly that this gas of eigenvalues is extremely strongly related to the ten dimensional geometry of the gravity dual. As a bonus, one also finds that extrapolating to strong coupling the off-diagonal modes connecting the eigenvalues are very massive (with a mass of the order of $\sqrt \lambda$) that is proportional to the distance between the eigenvalues. These modes can therefore be integrated out a la Born-Oppenheimer: they are fast degrees of freedom, while the eigenvalue dynamics has only the slow degrees of freedom. Not only did we find a round $S^5$, but we also found that the distance on the sphere becomes a measurement of masses, just like strings stretched between points on the sphere. In fact, this simple calculation, properly including the correct size of the sphere \cite{BCV} gives an exact agreement with string states moving on the $S^5$ \cite{HM} for strings of the size of the sphere. This calculation also matches the all-loop algebraic calculation of \cite{SZ}, where it is not clear how corrections to the approximations made should be included. 

This has been generalized to many other geometries $AdS_5\times X$, where the field theory duals are not weakly coupled, but where the eigenvalue  dynamics on the moduli space 
can still be studied within the same approximation \cite{BHart}. One needs to add by hand some postulate to compute the measure. Given these assumptions,  one matches a lot of non-trivial results in gravity: one finds that for both problems the spectral problem on $X$ that one has to solve is identical. The match of operator dimensions to supergravity masses is automatic, including supergarvity states that are not BPS protected. One also matches the plane wave limit calculation of the string spectrum and one can argue that one gets the correct metric on $X$. The case for three dimensional field theories is less well understood \cite{BT}, and again, there is no information on how to compute the function $h(\lambda)$ with these techniques. Agreement with gravity  requires corrections to the semiclassical measure that one computes in the four dimensional cases
and these feed into the calculation of $h(\lambda)$.

For me, personally, the fact that these techniques give some notion of approximate locality and higher dimensional metric geometry is rather satisfying. They also indicate how these notions of geometry can break down in the process of formation of black holes. The idea of how to build a black hole is simple: we put a lot of matter in some region and let it collapse. The matter here is going to be a dense gas of strings in some small region of the sphere. We turn this into an initial condition on the field theory that is given by turning on off-diagonal degrees of freedom between eigenvalues in some small region of the sphere. As we let the system evolve, the eigenvalues connected this way collapse and get trapped in some form of non-abelian thermal condensate that is well separated from the other degrees of freedom \cite{AspB}. One can give some account of why the black hole is black, thermal and why it evaporates slowly with negative specific heat, but precise details are beyond what one can calculate.  This way of thinking about the problem  has not yet answered questions of what are horizons and why do fluctuations in ten dimensions gravitate towards each other. 

In some sense, the best set of calculations that do address how gravitational interactions arise  form field theory dynamics predates  the modern version of AdS/CFT. These issues can be addressed for widely separated configurations of matter in the maximally supersymmetric quantum mechanics in $0+1$ dimensions \cite{BFSS} (see also \cite{DKPS}), which is claimed to be a description of M-theory on the lightcone frame. String interactions can be seen in the matrix string theory \cite{DVV}, which is described by a $1+1$ dimensional field theory living on a D1-brane.

The states dual to black holes can be also analyzed in these setups using lattice techniques in Euclidean Field Theory. One can match these to gravity calculations rather precisely. For recent work see \cite{Lattice}.

\section{Outlook}

So where are we now? 
Perhaps what is most important in this discussion is that even though over ten years have passed since the formulation of the AdS/CFT correspondence, we seem to still have a long way to go before we solve the most pressing  problems of quantum gravity from field theory, especially to address how holographic ideas get realized more locally on the geometry as well as to what happens to space-time when black holes form and evaporate.

Progress on understanding the duality has been incremental and steady. We have a remarkable list of exact solutions to field theory problems that match the weak gravity limit of various AdS/CFT setups, not only in four dimensions, but also in three dimensions. These give us hints on how to understand stringy and quantum corrections on the geometry in more general settings.  

The AdS/CFT correspondence seems to indicate that it is possible to reconcile gravity and quantum mechanics: Hawking's paradox is resolved. Albeit the solution is more akin to an existence proof of a solution than to a detailed understanding of how the information gets out of black holes and what they are made of.

It is encouraging that some of these problems might be solvable with numerics, as this might eventually give us data with which we can test our theories and understanding of geometry. In some sense the hardest obstacle to progressing in our understanding quantum gravity in general settings is that we have very little data to guide us.

\acknowledgments

Work supported in part by the DOE under grant DE-FG02-91ER40618.

\end{document}